\documentstyle[letters, twocolumn]{mn}
\def\beq{\begin{equation}}
\def\eeq{\end{equation}}
\def\bey{\begin{eqnarray}}
\def\eey{\end{eqnarray}}
\def\kms{\mbox{\rm \,km\,s}^{-1}}

\title[]
	{Possible Implications of Sagittarius Type Dwarf Galaxy on Microlensing Towards the Large Magellanic Cloud}
\author[]
	{HongSheng Zhao
	\thanks{has moved to Sterrewacht Leiden (hsz@strw.LeidenUniv.nl)} \\
	Max-Planck-Institute f\"ur Astrophysik,
Karl-Schwarzschild-Strasse 1, 85740 Garching, Germany}
\date{}
\pagerange{\pageref{firstpage}--\pageref{lastpage}}
\pubyear{1996}

\begin{document}
\maketitle
\label{firstpage}

\begin{abstract}

The discovery of the Sgr dwarf galaxy (Ibata, Gilmore, Irwin 1994)
shows that the Galactic halo contains large clumps which are not full
viralized.  Stars in such a clump can lense a background extragalactic
source.  I use parameters of Sgr to demonstrate that microlensing by a
disintegrated dwarf galaxy has a probability (optical depth)
$10^{-7}-10^{-6}$, which is comparable to that of an isothermal
Galactic dark halo of massive compact objects (MACHOs).  This may
change implications of observed microlensing events towards the LMC.
If more than one such clumps are hidden in the Galactic halo, then
given their enormous angular size ($\sim 5^o \times 20^o$), there is a
significant chance for {\sl any} line of sight, say the LMC direction,
to intersect one of them.  As an alternative to the half-MACHO dark
halo models (Alcock et al. 1996a,c), I discuss a non-MACHO model with
lenses of the observed events being faint stars in a hidden Sgr-like
disintegrated dwarf in the line-of-sight path to the LMC.

\end{abstract}


\section{Introduction}

A few dwarf galaxies are long known to orbit in the out-skirts of the
Galactic halo (Irwin and Hatzidimitriou 1995 and references therein).
But the recent discovery of a disintegrated dwarf galaxy at 24 kpc
from us in the Sagittarius constellation (Ibata, Gilmore, Irwin 1994,
1995) shows that some of them have spiraled {\sl in} the halo of the
Galaxy.  On the order of $10^{7-8} L_\odot$ luminosity of the Sgr
dwarf galaxy clusters in a narrow phase space (with roughly $10 \kms$
velocity dispersion) and is dispersed spatially in a large faint strip
probably along its orbit.  Several authors have identified parts of
dwarf galaxy across a large fraction of the sky (Mateo et al. 1996,
Alard 1996, Alcock et al. 1996b and references therein), and its
disintegrated material spans at least $5^o \times 20^o$ of the sky,
which is about $16$ kpc$^2$ in area.

It is surprising that such a ``big'' dwarf galaxy, which is also the
nearest and probably the most luminous Galactic dwarf galaxy, has
evaded previous frequent studies of the same region of the sky.  The
discovery naturally encourages one to speculate other dwarf galaxies
accreted by the Galaxy in the past one Hubble time (Mateo 1996); they
might have been torn into a strip similar to the Sgr dwarf by the
tidal force of the Galaxy, and might have been missed previously
because of the lack of an optimized systematic search.  On the other
hand if the halo hides several of these disintegrated dwarfs, each
occupying a significant solid angle of the sky, there is a fair chance
for one of them showing up in deep studies of an arbitrary line of
sight.

In this paper, I discuss how these hidden structures may affect the
ongoing microlensing search of MACHOs, massive compact dark matter in
the halo.  The microlensing probability (optical depth) of an
extragalactic source is determined by the density distribution of
lenses along the line of sight in the halo.  Most authors have used a
few observed microlensing events towards the LMC to draw conclusions
on the typical mass and the total mass of MACHOs (e.g. Alcock et
al. 1996a,c and references therein).  The results are based on the
assumption that lenses in the middle range of the line of sight to the
LMC are MACHOs which follow a smooth $r^{-2}$ law of a standard
isothermal halo.  Here I examine the effects of a clumpy halo due to
many disintegrated dwarfs in the halo.

\section{Lensing by faint stars in a Sgr-like dwarf galaxy}
\subsection{Optical depth}
In this section I compute the microlensing optical depth due to lenses
in a Sgr-like dwarf galaxy.  Lacking the knowledge of the real
distribution of these dwarf galaxies, I will simply ``move'' Sgr dwarf
galaxy in front of the LMC, and compute the microlensing probability
of source stars in the LMC.

The optical depth due to lenses in the dwarf galaxy is
\beq\label{dg}
\tau_{dg}={4 \pi G \over c^2} D \Upsilon \mu
= 0.17 \times 10^{-6} {\Upsilon \over 5 M_\odot L_\odot^{-1} } {\mu \over 4 L_\odot \mbox{\rm pc}^{-2} } {D \over 12 \mbox{\rm kpc} }
\eeq
where $D=D_d \left(1 - {D_d \over D_s}\right)$ is the effective
distance to the dwarf galaxy, given by $D_d$ and $D_s$ which are the
distances to the dwarf galaxy and the source.  The distance to the
Sgr dwarf galaxy is about ${1 \over 2}$ of the distance to the LMC, so
$D_s \approx 2 D_d \approx 50$ kpc, and $D \approx 12$ kpc.  $\mu$ is the
surface brightness of the dwarf galaxy, which for the Sgr 
is $4 L_\odot pc^{-2}$
near the nucleus and is decreased to about $1.5 L_\odot pc^{-2}$
at $10^o$ from the nucleus (Ibata et al. 1995, Mateo et al. 1995, 1996).
$\Upsilon$ is the ratio of total mass in stars and other compact
objects in the dwarf to its total stellar light.  Dwarf galaxies are
generally dominated by dark matter from the core to the tidal radius
with $\Upsilon$ probably in the range $5-200$ (Irwin and Hatzidimitriou 1995).

The observed optical depth to the LMC is still uncertain.  The number
of claimed microlensing events from MACHO and EROS experiments
has been fluctuating between 1 to
half a dozen.  Based on 6-8 events which pass their most recent
selection criteria the MACHO team 
estimated an optical depth towards the LMC
\beq \label{obs}
\tau_{obs} = 0.17 ^{+0.09}_{-0.05}\times 10^{-6};
\eeq
the most recent estimation (Alcock et al. 1996c) gives a somewhat higher value.

If one interprets these events as detection of MACHOs in an isothermal
Galactic dark halo, one would come to the conclusion that about
$30\%$ of the halo dark matter is in MACHOs.  A full MACHO halo would give
an optical depth 
\beq \label{dh}
\tau_{dh} \sim {V^2_{cir} \over c^2} \sim 0.5 \times 10^{-6},
\eeq
where $V_{cir} \sim 220 \kms$ is the amplitude of the gas rotation
curve of the Galaxy, insensitive to the line-of-sight direction.


Alternatively one can use objects in a foreground Sgr-like dwarf
galaxy as lenses.  In the case that the dark matter in both the
Galactic halo and dwarf galaxies is {\sl non-baryonic}, lenses can
only be {\sl faint stars of the dwarf}, which
depending on the shape of the mass
function at the lower mass end can have a mass-to-light
ratio $\Upsilon \approx (2-10)$.
So if Sgr dwarf galaxy were ``moved'' in front of the LMC,
then its faint stars can give an optical depth 
(cf. eq.~\ref{dg} and~\ref{obs})
\beq\label{comp}
\tau_{dg} = \left({\Upsilon \over 5 M_\odot L_\odot^{-1} }\right) \tau_{obs} = (0.4-2) \times \tau_{obs},~~~ \Upsilon \approx (2-10).
\eeq

It may appear surprising that faint stars in a $10^{8-9} M_\odot$ {\sl
disintegrated} dwarf galaxy can produce an optical depth comparable to
that of a $10^{12}M_\odot$ MACHO dark halo.  The optical depth is
roughly proportional to the projected density of and the distance to
the lenses (cf eq.~\ref{dg}).  A dwarf galaxy can be at a more
favorable lensing distance (roughly midway to the source stars) than
the MACHOs in the line of sight, which are in average at a distance
about $8$ kpc from us.  Also the projected mass density of a dwarf
galaxy, $\sim \left(4 \Upsilon\right) \sim 20 M_\odot pc^{-2}$, is
comparable to the MACHO halo, $\sim \left(V_{cir}^2 f_{MACHO}\right)
\left( 4 \pi G r\right)^{-1} \sim \left(100f_{MACHO}\right) \sim 30
M_\odot pc^{-2}$, where $f_{MACHO}$ is the fraction of the halo mass
in MACHOs.

The above shows that stars of a foreground Sgr-like dwarf galaxy can
in principle explain a significant fraction or all of the LMC
microlensing events.  This conclusion does {\sl not crucially} depend
on using the Sgr's surface brightness and distance at the present
epoch.  The optical depth reduces only by $20\%$ if the dwarf galaxy is at
$15$ or $35$ kpc, and by $40\%$ at $10$ or $40$ kpc.  The
dwarf's optical depth is proportional to its surface density
(cf. eq.~\ref{dg}), hence is a function of line-of-sight position and
age.  For the Sgr dwarf galaxy, the major axis gradient of the surface
density is shallow: I estimate an e-folding angular size from the
nucleus about $10^o$.  A dwarf with such size is big enough to
``cover'' the sky area of the LMC, in which case the optical depth is
insensitive to a small misalignment of the dwarf nucleus with the LMC,
and the microlensing events should not strongly cluster to the LMC
bar.\footnote{This feature is also shared by the isothermal halo
model, but not by models using stars in the LMC bar as lenses (Sahu,
1994).} Indeed, the MACHO team observed comparable numbers of events
towards the LMC bar and towards the fainter disc of the LMC.  Also the
optical depth of a disintegrating dwarf can even be bigger earlier on
when it is less spread out and is denser; comparing to the Sgr, an
undisrupted dwarf galaxy such as Fornax is much smaller and a factor
of a few denser in surface density (Irwin and Hatzidimitriou 1995).

If one relaxes the model assumption to allow many compact stellar
objects (such as stellar remnents and brown dwarfs) in both dwarf
galaxies and the Galactic halo, then the optical depth of the dwarf
can dominate that of the halo by as much as one order of magnitude
(cf. eq.~\ref{comp}).  This is because most of the gravitational mass
of a dwarf galaxy can be in compact objects with a mass-to-light ratio
$\Upsilon$ as large as $\sim 200M_{\odot}L_{\odot}^{-1}$.  Now suppose
that the Galactic halo is built up by accreting $10^{3-4}$
MACHO-dominated dwarfs gradually without going through a violent
relaxation phase.  Then particles of the dwarfs are probably not fully
mixed in the halo in a Hubble time, but are more likely only sheared
into many strips along their orbits, similar to the situation of the
Sgr dwarf.
\footnote{Such a distribution of dwarfs need not contradict the flat
rotation curve of gas at large radius as long as the mass enclosed
inside a radius satisfies the $M(r) \propto r$ law.}  Each dwarf might
cover, say a $5^o \times 20^o$ strip, which is $0.2\%$ of the full
sky.  Looking through this clumpy halo the optical depth will be a
wildly oscillating function of line-of-sight directions on angular
scale of dwarfs.  For example, an extragalactic source behind Sgr
could have a much larger optical depth than a source in a line of
sight offseted by a few degrees from Sgr.  In this picture one can not
argue any single line of sight, say the LMC, is typical with or
without a dwarf in front of the LMC.  It becomes problematic to
constrain the distribution of MACHOs with the observed microlensing
events in the LMC line-of-sight.

\subsection{Event time scales and lens mass function}

If the lenses of the LMC events were in a foreground dwarf galaxy rather the
isothermal MACHO halo, the implications on the lens's mass 
would be quite different.  The Einstein diameter crossing time $2t_E$ 
depends on the mass of the lens $m$, the relative transverse 
velocity between the lens and the source $V$ and the effective distance $D$
by the following simple equation,
\beq
2 t_E \approx 80 \mbox{\rm days} \left({m \over 0.3 M_\odot}\right)^{1 \over 2} \left({D \over 10 \mbox{\rm kpc}}\right)^{1 \over 2} \left({V \over 200 \kms}\right)^{-1}.
\eeq

The 6 events recently reported by the MACHO collaboration have
$2\times t_E=34-114$ days if excluding one binary event and one low
confidence event (Alcock et al. 1996c).  The MACHO team interpreted
these events by lenses in an isothermal halo, and suggested that
roughly half of the halo'mass is probably in locked up in $0.35
M_\odot$ white dwarfs.

Unlike the halo lenses which have a wide range of distances and
velocities, the lenses in a dwarf galaxy have negligible spread in
distance and velocity.  So the spread in the duration of observed
events is purely due to the mass function of the dwarf lenses.  The
roughly factor of $3$ range in the observed $t_E$ translates to about
a factor of $10$ in the lens mass $m$.  The actual median mass depends
sensitively on the proper motion of the dwarf galaxy.  But if I assume
that the dwarf galaxy has the average velocity of the halo stars, then
lens masses range between $0.1M_\odot$ to $1M_\odot$, which is
plausible for an old stellar population.

\section{Searching for disintegrated dwarf galaxies in the halo}

How tight are current observational constraints on any debris of a
disintegrated dwarf galaxy in front of the LMC?  Such a component
clearly has not been claimed in previous studies of LMC stellar
populations, but can be also difficult to rule out without an
optimized search of the neighboring sky region.

Stars in a dwarf galaxy are identifiable because they have a much
narrow spread in distance modulus, radial velocity, and proper motion
than field stars in the halo.  Metal poor RR Lyraes are by far the
best tracer.  The Sgr dwarf galaxy is detected by RR Lyraes in many
low extinction regions of the bulge, including the famous Baade window
(Ibata et al. 1994, Alard 1996, Alcock et al. 1996b).  Considering
that stellar populations of the Galactic bulge have been studied for
50 years since Baade's (1946) pioneering discovery of variable stars
in the direction, the delay of discovery is surprising and shows the
limitations of previous observations.  A distant disintegrated dwarf
galaxy can evade detection with a combination of large size and
distance, low surface density and any other foreground or background
extended structure.  The Sgr horizontal branch stars are spread out
over many degrees of the sky with a density about $1$ per square
arcmin (Ibata et al. 1995) and $\sim 19$ magnitude in $V$.  Their
faint extended structure will not stand out in a degree size
photographic plate, or a shallow color-magnitude diagram from a CCD
with only arcmin field of view.  Since roughly only one out of $30$ RR
Lyraes towards the bulge belongs to the dwarf galaxy (Alard 1996,
Alcock et al. 1996b), a large deep data set of RR Lyraes is necessary,
which also explains why the Sgr dwarf is discovered following the
recent large kinematic surveys and OGLE, DUO and MACHO microlensing
and variable star surveys towards the bulge.

The situation with the LMC is similar in many aspects except that (i)
the dwarf will be in the foreground rather than background, making its
RR Lyraes brighter than those in the LMC, and (ii) one is subject to
mistake an extended foreground structure as parts of the very
irregular structure of the LMC.  Curiously the Magellanic Clouds, the
Magellanic Stream, most of the known dwarf galaxies (Ursa Minor,
Draco, etc.) and several remote globular clusters, and a handful of
distant carbon stars (Irwin 1991 and references therein), all are
dispersed on either a great circle defined by the Magellanic Stream
(Lynden-Bell 1976) or a so-called Magellanic Plane (Kunkel 1979).  It
would be a natural speculation that an unobserved long stream of faint
stars, torn from an ancient Greater Magellanic Galaxy, might also lie
in such a plane with some debris in our line of sight to the LMC.

Payne-Gaposchkin (1971) published a survey of variable stars in the
direction of the LMC, listing 29 short-period variables as foreground
RR Lyraes with distances between 5 and 25 kpc.  Interestingly 9 of
these cluster at a narrow distance modulus range $16-17$ mag, which
corresponds to a line-of-sight distance about 16 to 25 kpc.  The
follow-up photometric and spectroscopic studies by Connolly (1985) and
Smith (1985) confirmed that most of these short-period variables are
indeed foreground metal poor RR Lyraes with a radial velocity
different from LMC RR Lyraes.  Since only about 1 RR Lyrae between 5
to 16 kpc, and also between 16 to 25 kpc towards the LMC are expected
from a smooth $r^{-3.5}$ or $r^{-4}$ density law for the RR Lyraes in
the stellar halo (Saha 1985), the 9 RR Lyraes with distance modulus
$16-17$ mag. seem to trace a local over-density region.  Whether there
is a genuine excess of RR Lyraes at half of the distance to the LMC
should be easily testable with large variable star data sets obtained
in microlensing experiments towards the LMC.


I thank Simon White for encouraging me in writing up this paper.  I am
obliged to a discussion with Mario Mateo and Lin Yan on observations.
I also thank Shude Mao, Ramesh Narayan and Hojun Mo for comments.

{}

\bsp
\label{lastpage}
\end{document}